\documentclass[12pt]{iopart}

\usepackage{graphicx}

\newcommand{\cd}{C$_6$D$_6$}
\newcommand{\cds}{C$_6$D$_6$ }

\newcommand{\lacls}{LaCl$_{3}$(Ce) }

\begin{document}

\title[Review and new concepts for neutron-capture measurements of astrophysical interest]{Review and new concepts for neutron-capture measurements of astrophysical interest}

\author{\footnotesize C. Domingo-Pardo$^{g,*}$, V. Babiano-Suarez$^{g}$, J. Balibrea-Correa$^{g}$, L. Caballero$^{g}$, I. Ladarescu$^{g}$, J. Lerendegui-Marco$^{r,g}$, J. L. Tain$^{g}$,  F. Calvi\~no$^{q}$, A. Casanovas$^{q}$, A. Segarra,$^{q}$
% ...
The n\_TOF Collaboration }

\centerline{\footnotesize$^{a}$ European Organization for Nuclear Research (CERN), Switzerland }
\centerline{\footnotesize$^{b}$ Centro de Investigaciones Energ\'eticas Medioambientales y Tecnol\'ogicas (CIEMAT),Spain }
\centerline{\footnotesize$^{c}$ INFN Laboratori Nazionali del Sud, Catania, Italy }
\centerline{\footnotesize$^{d}$ Dipartimento di Fisica e Astronomia, Universit\`a\ di Catania, Italy }
\centerline{\footnotesize$^{e}$ University of Lodz, Poland }
\centerline{\footnotesize$^{f}$ IPN, CNRS-IN2P3, Univ. Paris-Sud, Universit\'e\ Paris-Saclay, F-91406 Orsay Cedex,France }
\centerline{\footnotesize$^{g}$ Instituto de F\'\i sica Corpuscular, CSIC - Universidad de Valencia, Spain }
\centerline{\footnotesize$^{h}$ Technische Universit\"at Wien, Austria }
\centerline{\footnotesize$^{i}$ CEA Saclay, Irfu, Universit\'e\ Paris-Saclay, Gif-sur-Yvette, France }
\centerline{\footnotesize$^{j}$ Istituto Nazionale di Fisica Nucleare, Bari, Italy }
\centerline{\footnotesize$^{k}$ University of Manchester, United Kingdom }
\centerline{\footnotesize$^{l}$ Department of Physics, Faculty of Science, University of Zagreb, Croatia }
\centerline{\footnotesize$^{m}$ University of York, United Kingdom }
\centerline{\footnotesize$^{n}$ Istituto Nazionale di Fisica Nazionale, Perugia, Italy }
\centerline{\footnotesize$^{o}$ Dipartimento di Fisica e Geologia, Universit\`a\ di Perugia, Italy }
\centerline{\footnotesize$^{p}$ University of Santiago de Compostela, Spain }
\centerline{\footnotesize$^{q}$ Universitat Polit\`ecnica de Catalunya, Spain }
\centerline{\footnotesize$^{r}$ Universidad de Sevilla, Spain }
\centerline{\footnotesize$^{s}$ Istituto Nazionale di Astrofisica - Osservatorio Astronomico d'Abruzzo, Italy }
\centerline{\footnotesize$^{t}$ Dipartimento di Fisica, Universit\`a\ degli Studi di Bari, Italy }
\centerline{\footnotesize$^{u}$ National Technical University of Athens, Greece }
\centerline{\footnotesize$^{v}$ School of Physics and Astronomy, University of Edinburgh, United Kingdom }
\centerline{\footnotesize$^{w}$ Paul Scherrer Institut (PSI), Villigen, Switzerland }
\centerline{\footnotesize$^{x}$ Physikalisch-Technische Bundesanstalt (PTB), Bundesallee 100, 38116 Braunschweig, Germany }
\centerline{\footnotesize$^{y}$ University of Ioannina, Greece }
\centerline{\footnotesize$^{z}$ Instituto Superior T\'ecnico, Lisbon, Portugal }
\centerline{\footnotesize$^{aa}$ Joint Institute for Nuclear Research (JINR), Dubna, Russia}
\centerline{\footnotesize$^{ab}$ Goethe University Frankfurt, Germany }
\centerline{\footnotesize$^{ac}$ European Commission, Joint Research Centre, Geel, Retieseweg 111, B-2440 Geel,Belgium }
\centerline{\footnotesize$^{ad}$ Helmholtz-Zentrum Dresden-Rossendorf, Germany}
\centerline{\footnotesize$^{ae}$ Karlsruhe Institute of Technology, Campus North, IKP, 76021 Karlsruhe, Germany }
\centerline{\footnotesize$^{af}$ Japan Atomic Energy Agency (JAEA), Tokai-mura, Japan }
\centerline{\footnotesize$^{ag}$ Charles University, Prague, Czech Republic }
\centerline{\footnotesize$^{ah}$ Istituto Nazionale di Fisica Nucleare, Sezione di Bologna, Italy }
\centerline{\footnotesize$^{ai}$ Dipartimento di Fisica e Astronomia, Universit\`a\ di Bologna, Italy }
\centerline{\footnotesize$^{aj}$ Istituto Nazionale di Fisica Nucleare, Sezione di Legnaro, Italy }
\centerline{\footnotesize$^{ak}$ Consiglio Nazionale delle Ricerche, Bari, Italy}
\centerline{\footnotesize$^{al}$ Agenzia nazionale per le nuove tecnologie, l'energia e lo sviluppo economico sostenibile (ENEA), Bologna, Italy}
\centerline{\footnotesize$^{am}$ Istituto Nazionale di Fisica Nazionale, Trieste, Italy }
\centerline{\footnotesize$^{an}$ Horia Hulubei National Institute of Physics and Nuclear Engineering (IFIN-HH),Bucharest }
\centerline{\footnotesize$^{ao}$ University of Granada, Spain }
\centerline{\footnotesize$^{ap}$ University of Vienna, Faculty of Physics, Vienna, Austria }
\centerline{\footnotesize$^{aq}$ Bhabha Atomic Research Centre (BARC), India }
\centerline{\footnotesize$^{ar}$ Gran Sasso Science Institute (GSSI), L'Aquila, Italy}
\centerline{\footnotesize$^{as}$ Australian National University, Canberra, Australia }


\ead{domingo@ific.uv.es}
\vspace{10pt}
\begin{indented}
\item[]September 2019
\end{indented}

\begin{abstract}
The idea of slow-neutron capture nucleosynthesis formulated in 1957 triggered a tremendous experimental effort in different laboratories worldwide to measure the relevant nuclear physics input quantities, namely
($n,\gamma$) cross sections over the stellar temperature range (from few eV up to several hundred keV) for most of
the isotopes involved from Fe up to Bi. A brief historical review focused on total energy detectors will be presented to illustrate how,
advances in instrumentation have led, over the years, to the assessment and discovery of many new aspects of $s$-process
nucleosynthesis and to the progressive refinement of theoretical models of stellar evolution.
A summary will be presented on current efforts to develop new detection concepts, such as the Total-Energy Detector with $\gamma$-ray imaging capability (i-TED). The latter is based on the simultaneous combination of Compton imaging with neutron time-of-flight (TOF) techniques, in order to achieve a superior level of sensitivity and selectivity in the measurement of stellar neutron capture rates.
\end{abstract}

%
%
%
%
%

\section{Introduction}

Observation, theory and experiment: this seems to be a persistent repetitive pattern of research in nuclear astrophysics since its origin.
In 1952 Merrill observed for the first time absorption lines of the radioactive element Technecium in the stellar atmosphere of S-type stars~\cite{Merrill52}, hence revealing recent or ongoing nucleosynthesis activity in the stars. Five years later, the seminal theoretical works of Burbidge et al.~\cite{BBFH} and Cameron~\cite{Cameron57} (hereafter BBFHC) were published, thereby presenting a theory for the nucleosynthesis of heavy elements that essentially remains valid today~\cite{Kaeppeler11}. Shortly after BBFHC, a frenetic experimental activity followed in the nuclear physics laboratories, firstly intended to demonstrate and secondly to probe and constrain the predictions of the models and the observations by the astronomers. This contribution focuses on some of those experimental efforts. Sec.~\ref{sec:developments} describes a few examples selected to illustrate how advances in instrumentation and new detection concepts have led, over the years, to the progressive refinement of theoretical models and to a better understanding of the physical conditions in the corresponding stellar environments. Ongoing research to enhance further the sensitivity of TOF measurements and access more challenging measurements is presented in Sec.~\ref{sec:ited}.

\section{Instrumental developments and discoveries}\label{sec:developments}

There exist different techniques to determine neutron capture cross sections, that are the key nuclear physics input quantities for s-process model calculations. One of them is the activation method, which shows an extremely high sensitivity and selectivity (for recent examples see e.g.~\cite{Reifarth03,Uberseder09}). However, the activation method is mostly applicable to nuclei that, upon capture, become radioactive and have a convenient half-life and decay pattern. In some cases, the use of Accelerator Mass Spectrometry has allowed to overcome this limitation~\cite{Wallner07,Wallner12}.

\begin{figure}[!htbp]
    \begin{center}
      \includegraphics[width=0.8\columnwidth]{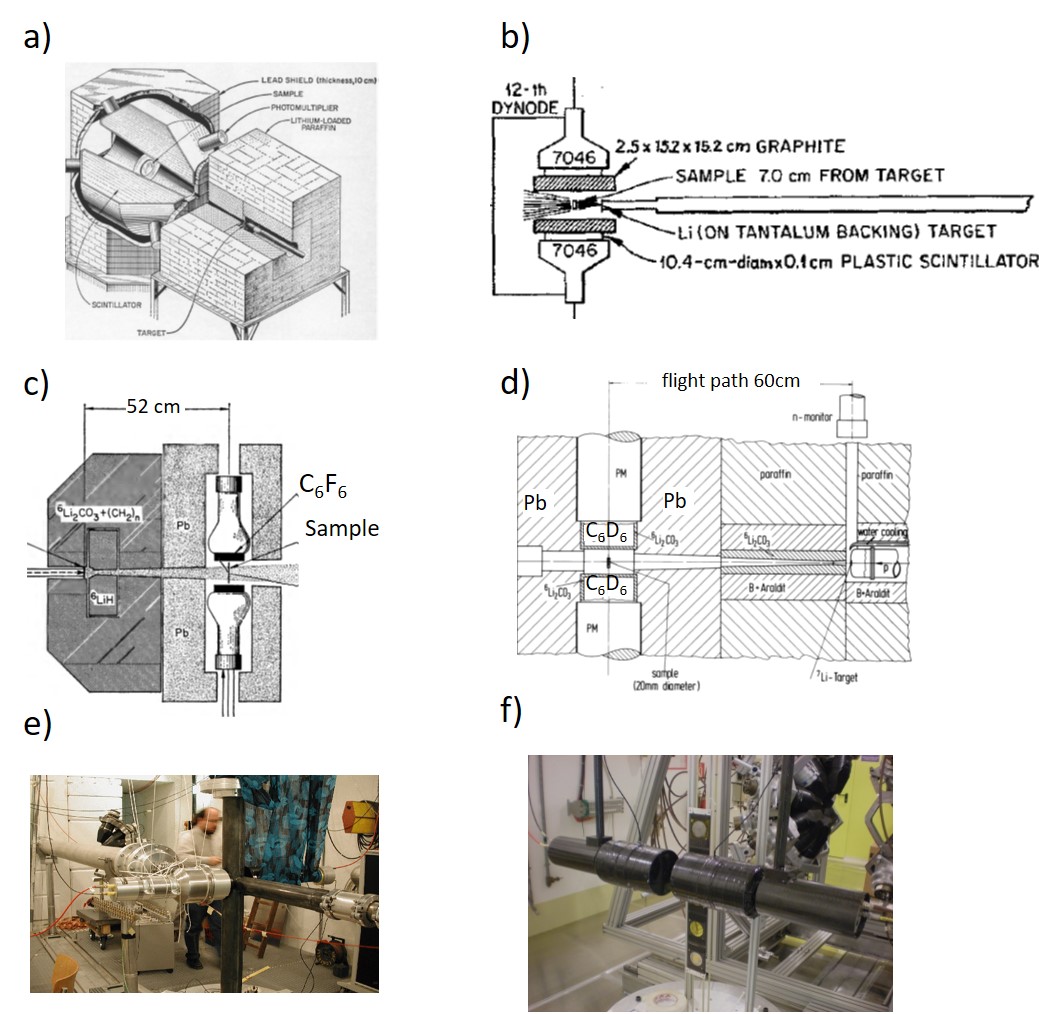}
      \end{center}
\caption{\label{fig:setups} a) Scintillation tanks used for TOF experiments by the time when BBFHC was published~\cite{Gibbons61}. b) Moxon-Rae detector developed for the first measurement of samples enriched on Sn and Sm isotopes (adapted from~\cite{Macklin63c}), applied to the first experimental confirmation of the s-process theory~\cite{Macklin62}. c) C$_6$F$_6$ setup used with the PHWT to measure $^{99}$Tc(n,$\gamma$) and to determine the AGB lifetime in the 3$^{rd}$ dredge-up phase~\cite{Winters87}. d) Neutron sensitivity improved set-up with C$_6$D$_6$ detectors~\cite{Walter86}. e) Unshielded low-background C$_6$D$_6$ detectors used for the first $^{151}$Sm(n,$\gamma$) TOF measurement~\cite{Abbondanno04}. f) Ultra-low neutron-sensitivity set-up similar as the one used for the $^{63}$Ni(n,$\gamma$) TOF experiment to constrain pre-supernova Cu-content in massive stars~\cite{Lederer13}.}
\end{figure}

In this contribution we focus on the use of pulsed neutron beams in combination with the time-of-flight (TOF) technique to measure neutron capture rates of astrophysical interest. The TOF method is applicable to any nucleus, provided that a sufficiently large sample of material becomes available. Also, it offers the possibility to cover, in a single measurement, the full stellar energy range of interest. By the time when the nucleosynthesis theory of BBFHC was published neutron capture TOF measurements were mostly based on the use of large scintillation tanks (1000 liters)~\cite{Gibbons61} with a large $\gamma$-ray detection efficiency, as the one shown in Fig.\ref{fig:setups}-a). However, the bulky setup showed a high sensitivity to neutron induced backgrounds, which strongly limited the detection sensitivity for the capture channel of interest. Indeed, measurements were restricted to only very large samples ($\sim$1~mole). Isotopically enriched samples of that size were very expensive and difficult to produce. Thus, the technique was mainly used to measure samples of elements with natural isotopic abundances~\cite{Gibbons61}. As suggested by BBFHC, in order to quantitatively test the s-process theory it was necessary to know the abundances and the cross-sections of the individual involved isotopes, as the relative elemental abundances may be altered by uncertain physical or chemical fractionation processes.

The first breakthrough in the field came with the development and application of a radically new concept: radiation detectors that featured a detection efficiency proportional to the $\gamma$-ray energy, so-called Moxon-Rae detectors~\cite{Moxon63,Macklin63a}, shown in Fig.~\ref{fig:setups}-b). These detectors were based on three layers of graphite, bismuth and a plastic scintillator. The latter was readout with a photomultiplier tube. The two front layers of C and Bi acted as electron converters for the incident radiation, thereby yielding a $\gamma$-ray detection efficiency roughly proportional to its initial energy. This proportionality condition between efficiency and $\gamma$-ray energy was of pivotal importance in order to avoid systematic errors in the cross section related to the multiplicity $m$ of the capture cascade or to the particular resonance decay path. In short, by ensuring a sufficiently low detection efficiency $i)\,\varepsilon_{\gamma}<<$, and the  proportionality condition $ii)  \varepsilon_{\gamma}=kE_{\gamma,j}$, the probability to detect a capture event $\varepsilon_c$ (eq.~\ref{eq:Ec}) becomes proportional to the capture energy, $\alpha E_c$, which is a constant well-defined value for each measured neutron energy ($E_c = S_n + E_n$).

\begin{equation}\label{eq:Ec}
\varepsilon_c = 1 - \prod^{m}_{j=1} (1 - \varepsilon_{\gamma,j}) \stackrel{i) }{\simeq} \sum^{m}_{j=1} \varepsilon_{\gamma,j} \stackrel{ii)}{\simeq} \alpha \sum^{m}_{j=1} \varepsilon_{\gamma,j} = \alpha E_c.
\end{equation}

\begin{figure}[!htbp]
  \begin{center}
\includegraphics[width=0.8\columnwidth]{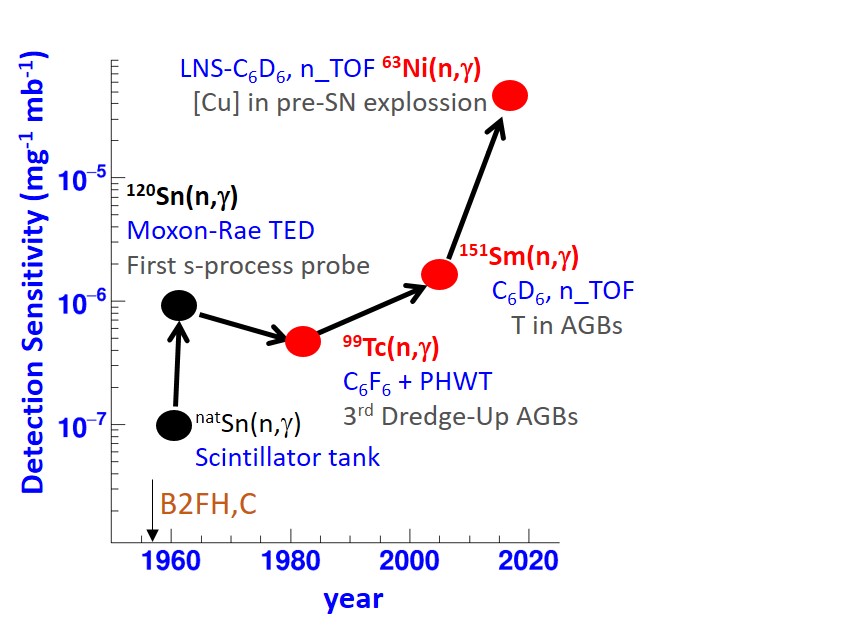}
  \end{center}
\caption{\label{fig:sensitivity} This figure shows some selected examples to illustrate how advances in the instrumentation helped first to validate, and afterwards assess and constrain many fascinating aspects of the stellar evolution. Sample radioactivity is not included in the sensitivity definition, but measurements with radioactive samples are shown with bold red circles. See text for details.}
\end{figure}

Moxon-Rae detectors, also called Total Energy Detectors (TEDs), attained about one order of magnitude improvement in detection sensitivity, thereby enabling measurements on samples ten times smaller ($\sim$0.1~mole) than those needed with the large scintillation tanks. Such sample quantities were better suited for isotopic enrichment and thus, the first measurement on isotopically enriched Tin isotopes could be carried out using samples of ``only'' 30-35~g~\cite{Macklin62}. Fig.~\ref{fig:sensitivity} schematically shows the evolution in time of the capture detection sensitivity using the TOF technique with low-volume detectors. The birth of the modern nucleosynthesis theory is indicated by the BBFHC label. Detection sensitivity is arbitrarily defined here as the inverse of the sample mass (mg) times the cross section (mb) at neutron energies of $kT\sim$30~keV. Thus, measurements of smaller sample quantities or lower cross sections indicate a higher capability of the detection apparatus to deliver results. Note that the examples shown in this contribution are not necessarily representative of the state-of-the-art at that time. Instead, they have been rather chosen in order to discuss the evolution of the methods in the field. It is also worth to emphasize, that enhancements in detection sensitivity have been naturally due to concomitant improvements in accelerators, neutron-beams and sample purification and preparation techniques. However, these aspects will not be discussed further in this contribution.

The measurement of the tin nuclei~\cite{Macklin62} was shortly followed by another measurement on isotopically enriched Sm samples~\cite{Macklin63c}, both of them using Moxon-Rae TEDs. At that time, these results represented a first experimental validation of the s-process theory, because they demonstrated the inverse proportionality between the nuclear neutron capture cross section and the relative (s-process) isotopic abundances, as predicted by the s-process theory of BBFHC~\cite{Macklin62}. Quoting~\cite{Macklin63c}, \emph{the results thus strongly confirm the s-process nucleosynthesis prediction}.

Despite their remarkable performance, the efficiency of Moxon-Rae detectors was so small, that very short flight-paths of only 5-10~cm could be utilized with the neutron beam intensities available at that time~\cite{Macklin63a}. However, the efficiency-energy proportionality embedded within the Moxon-Rae detectors was a very powerful concept, that would be highly exploited in the following decades.

The next notable development was to realize, that the proportionality condition could be also achieved by using almost any type of fast radiation detector of low neutron sensitivity in combination with a pulse-height analyzer and a recording system. By storing the amplitude of the pulses during the measurement a convenient ``weight'' could be applied afterwards in order to recover the proportionality condition, without need of (hardware) photo-electron converters. This was the origin of the so-called Pulse-Height Weighting Technique (PHWT)~\cite{Macklin67}. Large volumes ($\sim$0.6~L) of non-hydrogenous scintillation liquid (C$_6$F$_6$) were readout with photomultiplier tubes~\cite{Macklin67}, as shown in Fig.~\ref{fig:setups}-c). The higher efficiency of this new technique allowed one to use longer flight-paths of about 50~cm, and also to tackle the measurement of more challenging and smaller samples, such as radioactive samples of $^{99}$Tc~\cite{Winters87} with a mass of only 2.3~g ($\sigma^{99Tc}_{25~keV} = 933$~mb see Fig.~\ref{fig:sensitivity}). The decrease of sensitivity in Fig.~\ref{fig:sensitivity} for this example is only apparent, due to the simplified definition used here for detection sensitivity, which does not account for the experimental difficulty associated with the radioactivity of the sample itself. The cross-section measurement of $^{99}$Tc, in combination with Tc, Nb and Mo abundances observed in the stellar atmospheres of two S-type stars (R CMi and CY Cyg) allowed for the first determination of the lifetime spent by these stars in the 3$^{rd}$ dredge-up phase of evolution~\cite{Mathews86,Winters87}. Quite a remarkable result.

Further efforts to improve detection sensitivity focused on the reduction of the intrinsic neutron sensitivity in the detectors themselves, thereby replacing the fluorine in the liquid scintillator (C$_6$F$_6$) by deuterium (C$_6$D$_6$)~\cite{Walter86,Corvi88}, as shown in Fig.~\ref{fig:setups}-d). Later, it was found that the massive lead shielding around the scintillation detectors (Fig.~\ref{fig:setups}-c)-d)) was, in most cases, amplifying the background rather than suppressing it. New lightweight and unshielded C$_6$D$_6$-based detection systems, as the one shown in Fig.~\ref{fig:setups}-e), enabled many new astrophysical relevant results. One example is the first TOF measurement of the $s$-process branching nucleus $^{151}$Sm(n,$\gamma$) using only $\sim$200~mg of sample mass~\cite{Abbondanno04} ($\sigma^{151Sm}_{25~keV} = 3031$~mb, see Fig.~\ref{fig:sensitivity}). Under the He-burning conditions of AGB stars the $\beta$-decay rate of $^{151}$Sm ($t_{1/2}$=93~y) is significantly enhanced due to the thermal population of low-lying excited states. This effect can be used as a potential $s$-process thermometer. The measured cross section for $^{151}$Sm(n,$\gamma$), in combination with stellar models and the solar system abundance of $^{152}$Gd, allowed to constrain the temperature range of the He-shell flashes to 2.5-2.8$\times$10$^8$~K.

Presently, the state-of-the-art in TEDs is represented by C-fibre optimized detectors~\cite{Plag03} and surrounding structural elements~\cite{Marrone04} (Fig.~\ref{fig:setups}-f). The aim is to minimize the overall intrinsic neutron sensitivity of the detection set-up. Also, large volumes of $\sim$1~L are possible without necessarily compromising the applicability and accuracy of the PHWT, despite of the ineluctable $\gamma$-ray summing in each detector. For this, a sophisticated methodology based on MC-simulations and the statistical nuclear model allow to reliably account for the enhanced $\gamma$-ray summing effect, permitting an overall accuracy of better than 2\%~\cite{Tain04}. The capture detection sensitivity achieved with these low neutron sensitivity (LNS) \cds can be exemplified by the (n,$\gamma$) measurement of the unstable $^{63}$Ni(t$_{1/2}$=101.2(15)~y)~\cite{Lederer13}. The latter, in conjunction with stellar models for 25~M$_{\odot}$~\cite{Pignatari10}, could be used to constrain the Cu-composition of the s-process inventory in massive stars at their last evolutionary stage before exploding as supernovae~\cite{Lederer13}.

\section{Current efforts: $\gamma$-ray vision for background discrimination}\label{sec:ited}

One of the limiting factors in current TOF experiments with state-of-the-art detectors is due to neutrons scattered in the capture-sample, and subsequently thermalized and captured in the surrounding walls of the experimental area. After capture in the walls, these stray neutrons emit radiation which eventually reaches the \cd, thus enhancing the background level. This type of background is described in detail in Ref.~\cite{Zugec14} and actually limits the attainable peak-to-background ratio or detection sensitivity in many experiments already beyond few keV of neutron energy. One example is the measurement of $^{93}$Zr(n,$\gamma$)~\cite{Tagliente13,Neyskens15}, where capture levels beyond $E_n\sim 8$~keV were already difficult to identify. The spectrum labeled as ``Setup'' in Fig.~1 of Ref.~\cite{Tagliente13} shows the impact of this type of background, and reflects the limitation of existing systems to measure the cross section in the full energy range of stellar relevance. 

One possibility to reduce this type of background would be to ``focus'' the \cds towards the sample and shield them from the surroundings, so that they can only ``see'' true capture $\gamma$-rays coming from the sample. This is not a new idea and, indeed, the first experimental setups using TEDs with the PHWT were already utilizing some sort of mechanical collimation, as it is demonstrated in Fig.~\ref{fig:setups}-c) and d). However, as discussed in the previous section, the massive lead shielding is only effective for the suppression of $\gamma$-ray backgrounds, but not for the sample-scattered neutrons which produce further radiation in lead. This was also confirmed recently by using a $\gamma$-camera at CERN n\_TOF, which consisted of a position-sensitive radiation detector coupled to a heavy pin-hole collimator made from lead~\cite{Perez16}.

\begin{figure}[!htbp]
  \begin{center}
    \includegraphics[width=0.28\columnwidth]{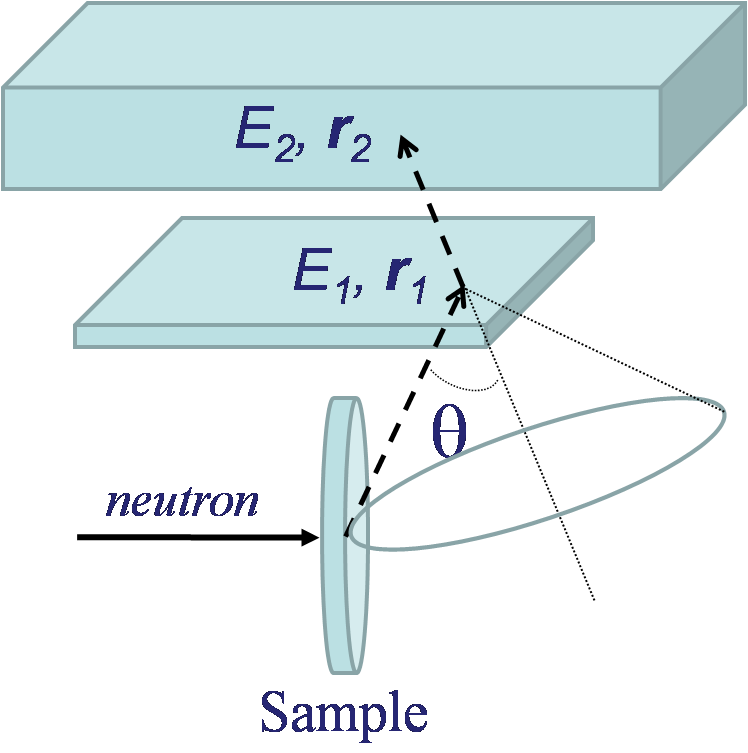}
\hfill
    \includegraphics[width=0.45\columnwidth]{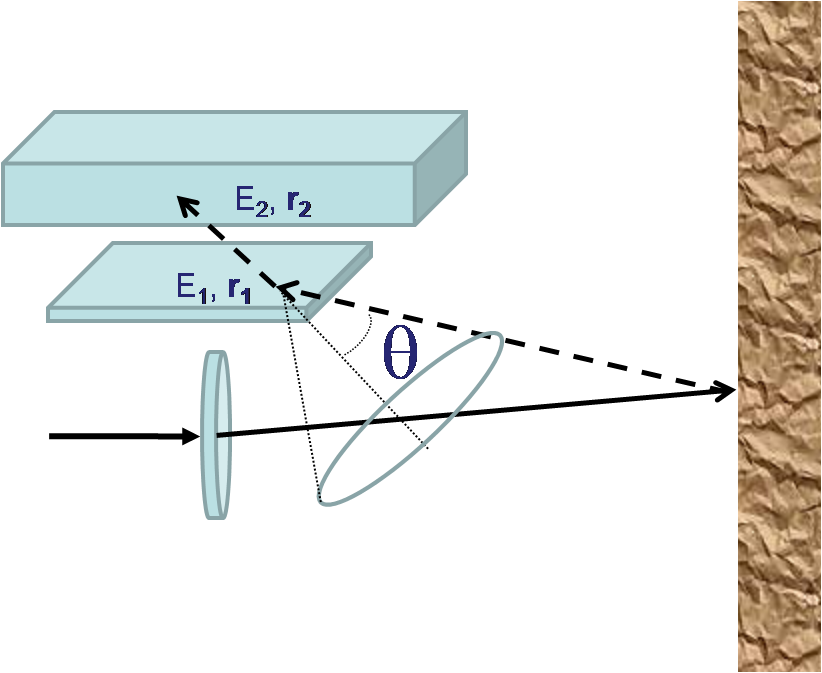}
  \end{center}
\caption{\label{fig:ited_concept} Schematic representation of the i-TED concept to discriminate true capture events (left) from contaminant radiation originating in the surroundings (right).} 
\end{figure}

Another possibility to ``focus'' the detectors towards the sample, without the need of mechanical collimation consists of using electronic collimation. Electronic collimation is based on the use of the Compton scattering law in order to infer a cone of possible incident directions~\cite{Schoenfelder73}. For this to be implemented one needs a detection system, normally divided in two volumes, that can provide information on the position and the energy of the $\gamma$-ray interactions in the detector (see Fig.~\ref{fig:ited_concept}). Further, for neutron capture TOF experiments the detection system must show a fast response and low sensitivity to neutrons. These are the main goals for a new detection system called i-TED (Total Energy Detector with $\gamma$-ray imaging capability)~\cite{Domingo16} that is being developed in the framework of the HYMNS project~\cite{hymns}. The proposed Compton imager is quite different from any previous detector used for (n,$\gamma$) measurements, a fact which exemplifies the versatility of the PHWT discussed in Sec.~\ref{sec:developments} where only the requirements of low-efficiency and efficiency-energy proportionality are required to reliably measure a cross section. Low efficiency is guaranteed in i-TED because the time-coincidence between scatter- and absorber-planes already reduces the overall efficiency a lot. The applicability of the PHWT to achieve the proportionality condition has been also demonstrated on the basis of MC simulations~\cite{Domingo16}.

\begin{figure}[!htbp]
  \begin{center}
    \includegraphics[width=0.48\columnwidth]{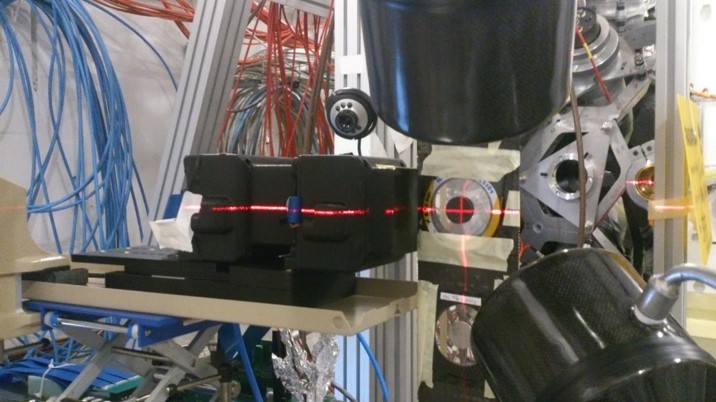}
    \includegraphics[width=0.43\columnwidth]{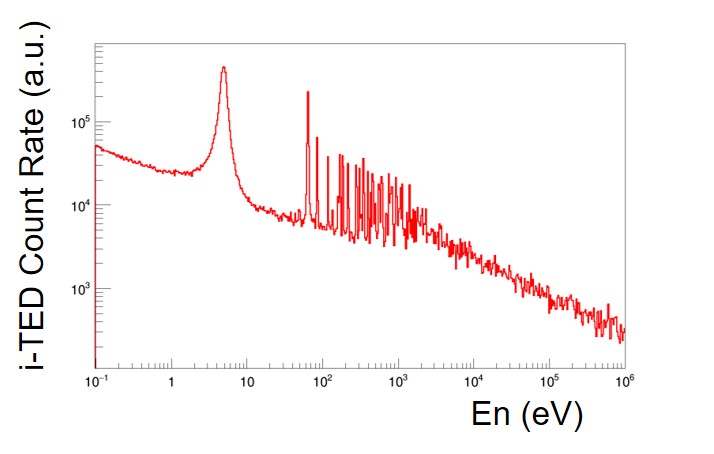}
  \end{center}
\caption{\label{fig:ited_au} (Left) Picture of the experimental set-up at CERN n\_TOF with an i-TED prototype being aligned at $\sim$5~cm from the capture sample. (Right) Un-weighted count-rate histogram as a function of the neutron energy measured with i-TED.}
\end{figure}

In order to attain sufficient angular resolution ($< 15^{\circ}$) for background rejection, high-resolution inorganic LaCl$_3$(Ce) scintillators are being implemented in i-TED~\cite{Babiano19b}. High-resolution HPGe~\cite{Mihailescu07} or other semiconductor sensors are excluded due to their slow time-response, their higher neutron capture cross section and their lower intrinsic efficiency. Energy- and position-information from the \lacls is gained by optically coupling them to pixelated silicon photomultipliers (SiPMs). In particular, we use $50\times 50 \times 10$~mm$^3$ monolithic crystals for the i-TED scatter plane and 2.5 thicker crystals for the absorber detection plane. Compared to other similar Compton cameras developed for medical applications~\cite{Kishimoto17,Nagao18}, i-TED uses significantly larger crystals which represents a challenge in terms of readout channels and overall performance. However, this is required for the intended TOF capture experiments in order to have a detection efficiency which is sufficient to perform a neutron capture experiment in a reasonable time frame.
The energy resolution that has been achieved with this type of position-sensitive detectors (PSDs) using SiPMs is of about 4-5\% \textsc{fwhm} at 662~keV~\cite{Olleros18}. The spatial resolution attainable with large monolithic crystals is also a technically difficult aspect. Thus far, we have investigated different approaches to reconstruct the 3D-coordinates of the $\gamma$-ray hit location in the \lacls volume. Presently, the most suited method seems to be a least-squares fit on an event-by-event basis to the SiPM response using an analytic model~\cite{Li10}, which accounts for the spatial propagation and reflection of the scintillation light produced by a point-like source in the crystal volume. By means of an accurate spatial calibration of our PSDs using a collimated $\gamma$-ray source in conjunction with an XY-positioning table, we have been able to obtain position resolutions of 1-2~mm \textsc{fwhm} at 511~keV for crystal thicknesses between 10~mm and 30~mm~\cite{Babiano19}.
As shown in Ref.~\cite{Domingo16}, the full i-TED system will comprise four large solid-angle Compton cameras arranged in a close geometry around the capture sample. In order to instrument the full array, about 1280~readout channels will be required. The SiPM signals are acquired using ASIC-based frontend readout electronics that have been developed by PETsys Electronics~\cite{petsys} for medical applications. The suitability of this readout approach for i-TED resides on its scalability and on the fact that it provides both energy (QDC) and time information with good resolution~\cite{Olleros18,Babiano19}. In order to adapt the medical readout system to the neutron-TOF requirements, some customization of the electronics was needed. In particular an external trigger-signal from the CERN-PS was fed and time-stamped in one of the readout-channels in order to enable the possibility to build a TOF spectrum. A first i-TED prototype has been already tested at the CERN n\_TOF EAR1 facility, and a count-rate versus neutron-energy spectrum is shown in Fig.~\ref{fig:ited_au} for $^{197}$Au(n,$\gamma$). The latter represents a technical validation of the system in terms of TOF performance, as no degradation of the neutron-energy resolution is observed with the use of this new instrumentation.

Once the system was adapted for TOF experiments and technically validated, further capture tests were carried out at CERN n\_TOF using an i-TED prototype more similar to the one described in Ref.~\cite{Babiano19b}. The aim of these latter tests was to explore the applicability of the $\gamma$-ray imaging for background rejection. The data are being analyzed and results will be reported in future publications.

\section{Summary and outlook}
Several neutron capture experiments and the related instrumentation have been discussed in order to illustrate how progress in capture detection systems has helped, since the origin of the modern nucleosynthesis theory, to validate and constrain different astrophysical aspects. The emphasis here has been put on the working principle of the so-called Total Energy Detectors (TEDs) and their interesting evolution along the nearly 70~years of existence. Finally, a new type of TED based on Compton imaging has been presented as a possibility to further reduce neutron-induced gamma-ray backgrounds and thus enhance the signal-to-background ratio in this type of experiments.

\section{Acknowledgments}
This work has received funding from the European Research Council (ERC) under the European Union's Horizon 2020 research and innovation programme (ERC Consolidator Grant project HYMNS, with grant agreement n$^{\circ}$ 681740). The authors acknowledge support from the Spanish Ministerio de Ciencia e Innovaci\'on under grants FPA2014-52823-C2-1-P, FPA2017-83946-C2-1-P, CSIC for funding PIE-201750I26 and the program Severo Ochoa (SEV-2014-0398). 

\section*{References}
\bibliography{bibliography}

\begin{thebibliography}{10}

\bibitem{Merrill52}
Paul~W. {Merrill}.
\newblock {Spectroscopic Observations of Stars of Class}.
\newblock {\em The Astroph. Jour.}, 116:21, Jul 1952.

\bibitem{BBFH}
E.~Margaret Burbidge, G.~R. Burbidge, William~A. Fowler, and F.~Hoyle.
\newblock Synthesis of the elements in stars.
\newblock {\em Rev. Mod. Phys.}, 29:547--650, Oct 1957.

\bibitem{Cameron57}
A.~G.~W. {Cameron}.
\newblock {On the origin of the heavy elements.}
\newblock {\em Astronomical Journal}, 62:9--10, Feb 1957.

\bibitem{Kaeppeler11}
F.~{K{\"a}ppeler}, R.~{Gallino}, S.~{Bisterzo}, and Wako {Aoki}.
\newblock {The s process: Nuclear physics, stellar models, and observations}.
\newblock {\em Reviews of Modern Physics}, 83(1):157--194, Jan 2011.

\bibitem{Reifarth03}
R.~Reifarth, C.~Arlandini, M.~Heil, F.~Kappeler, P.~V. Sedyshev, A.~Mengoni,
  M.~Herman, T.~Rauscher, R.~Gallino, and C.~Travaglio.
\newblock Stellar neutron capture on promethium: Implications for thes-process
  neutron density.
\newblock {\em The Astrophysical Journal}, 582(2):1251--1262, jan 2003.

\bibitem{Uberseder09}
E.~Uberseder, R.~Reifarth, D.~Schumann, I.~Dillmann, C.~Domingo Pardo,
  J.~G\"orres, M.~Heil, F.~K\"appeler, J.~Marganiec, J.~Neuhausen,
  M.~Pignatari, F.~Voss, S.~Walter, and M.~Wiescher.
\newblock Measurement of the
  $^{60}\mathrm{Fe}(n,\ensuremath{\gamma}{)}^{61}\mathrm{Fe}$ cross section at
  stellar temperatures.
\newblock {\em Phys. Rev. Lett.}, 102:151101, Apr 2009.

\bibitem{Wallner07}
Anton {Wallner}, Max {Bichler}, Iris {Dillmann}, Robin {Golser}, Franz
  {K{\"a}ppeler}, Walter {Kutschera}, Michael {Paul}, Alfred {Priller}, Peter
  {Steier}, and Christof {Vockenhuber}.
\newblock {AMS measurements of $^{41}$Ca and $^{55}$Fe at VERA two
  radionuclides of astrophysical interest}.
\newblock {\em Nuclear Instruments and Methods in Physics Research B},
  259(1):677--682, Jun 2007.

\bibitem{Wallner12}
A.~{Wallner}, K.~{Buczak}, I.~{Dillmann}, J.~{Feige}, F.~{K{\"a}ppeler},
  G.~{Korschinek}, C.~{Lederer}, A.~{Mengoni}, U.~{Ott}, M.~{Paul},
  G.~{Sch{\"a}tzel}, P.~{Steier}, and H.~P. {Trautvetter}.
\newblock {AMS Applications in Nuclear Astrophysics: New Results for
  $^{13}$C(n,{\ensuremath{\gamma}})$^{14}$C and $^{14}$N(n,p)$^{14}$C}.
\newblock {\em Publications of the Astronomical Society of Australia},
  29(2):115--120, May 2012.

\bibitem{Gibbons61}
J.~H. {Gibbons}, R.~L. {Macklin}, P.~D. {Miller}, and J.~H. {Neiler}.
\newblock {Average Radiative Capture Cross Sections for 7- to 170-kev
  Neutrons}.
\newblock {\em Physical Review}, 122(1):182--201, Apr 1961.

\bibitem{Macklin63c}
R.~L. {Macklin}, J.~H. {Gibbons}, and T.~{Inada}.
\newblock {Neutron Capture in the Samarium Isotopes and the Formation of the
  Elements of the Solar System}.
\newblock {\em Nature}, 197(4865):369--370, Jan 1963.

\bibitem{Macklin62}
R.~L. {Macklin}, T.~{Inada}, and J.~H. {Gibbons}.
\newblock {Neutron Capture in Tin Isotopes at Stellar Temperatures}.
\newblock {\em Nature}, 194(4835):1272, Jun 1962.

\bibitem{Winters87}
R.~R. {Winters} and R.~L. {Macklin}.
\newblock {Maxwellian-averaged Neutron Capture Cross Sections for 99Tc and
  95-98Mo}.
\newblock {\em The Astroph. Jour.}, 313:808, Feb 1987.

\bibitem{Walter86}
G.~{Walter}, H.~{Beer}, F.~{Kaeppeler}, G.~{Reffo}, and F.~{Fabbri}.
\newblock {The s-process branching at Se-79}.
\newblock {\em Astronomy and Astrophysics}, 167(1):186--199, Oct 1986.

\bibitem{Abbondanno04}
U.~Abbondanno, G.~Aerts, F.~Alvarez-Velarde, H.~\'Alvarez-Pol, S.~Andriamonje,
  J.~Andrzejewski, G.~Badurek, P.~Baumann, F.~Be\ifmmode \check{c}\else
  \v{c}\fi{}v\'a\ifmmode~\check{r}\else \v{r}\fi{}, J.~Benlliure,
  E.~Berthoumieux, F.~Calvi\~no, D.~Cano-Ott, R.~Capote, P.~Cennini, V.~Chepel,
  E.~Chiaveri, N.~Colonna, G.~Cortes, D.~Cortina, A.~Couture, J.~Cox,
  S.~Dababneh, M.~Dahlfors, S.~David, R.~Dolfini, C.~Domingo-Pardo, I.~Duran,
  M.~Embid-Segura, L.~Ferrant, A.~Ferrari, R.~Ferreira-Marques,
  H.~Frais-Koelbl, W.~Furman, I.~Goncalves, R.~Gallino, E.~Gonzalez-Romero,
  A.~Goverdovski, F.~Gramegna, E.~Griesmayer, F.~Gunsing, B.~Haas, R.~Haight,
  M.~Heil, A.~Herrera-Martinez, S.~Isaev, E.~Jericha, F.~K\"appeler, Y.~Kadi,
  D.~Karadimos, M.~Kerveno, V.~Ketlerov, P.~Koehler, V.~Konovalov,
  M.~Krti\ifmmode~\check{c}\else \v{c}\fi{}ka, C.~Lamboudis, H.~Leeb,
  A.~Lindote, I.~Lopes, M.~Lozano, S.~Lukic, J.~Marganiec, S.~Marrone,
  J.~Martinez-Val, P.~Mastinu, A.~Mengoni, P.~M. Milazzo, A.~Molina-Coballes,
  C.~Moreau, M.~Mosconi, F.~Neves, H.~Oberhummer, S.~O'Brien, J.~Pancin,
  T.~Papaevangelou, C.~Paradela, A.~Pavlik, P.~Pavlopoulos, J.~M. Perlado,
  L.~Perrot, M.~Pignatari, R.~Plag, A.~Plompen, A.~Plukis, A.~Poch,
  A.~Policarpo, C.~Pretel, J.~Quesada, S.~Raman, W.~Rapp, T.~Rauscher,
  R.~Reifarth, M.~Rosetti, C.~Rubbia, G.~Rudolf, P.~Rullhusen, J.~Salgado,
  J.~C. Soares, C.~Stephan, G.~Tagliente, J.~Tain, L.~Tassan-Got, L.~Tavora,
  R.~Terlizzi, G.~Vannini, P.~Vaz, A.~Ventura, D.~Villamarin, M.~C. Vincente,
  V.~Vlachoudis, F.~Voss, H.~Wendler, M.~Wiescher, and K.~Wisshak.
\newblock Neutron capture cross section measurement of
  $^{151}\mathrm{S}\mathrm{m}$ at the cern neutron time of flight facility
  (n\_tof).
\newblock {\em Phys. Rev. Lett.}, 93:161103, Oct 2004.

\bibitem{Lederer13}
C.~{Lederer}, C.~{Massimi}, S.~{Altstadt}, J.~{Andrzejewski}, L.~{Audouin},
  M.~{Barbagallo}, V.~{B{\'e}cares}, F.~{Be{\v{c}}v{\'a}{\v{r}}}, F.~{Belloni},
  E.~{Berthoumieux}, J.~{Billowes}, V.~{Boccone}, D.~{Bosnar}, M.~{Brugger},
  M.~{Calviani}, F.~{Calvi{\~n}o}, D.~{Cano-Ott}, C.~{Carrapi{\c{c}}o},
  F.~{Cerutti}, E.~{Chiaveri}, M.~{Chin}, N.~{Colonna}, G.~{Cort{\'e}s}, M.~A.
  {Cort{\'e}s-Giraldo}, M.~{Diakaki}, C.~{Domingo-Pardo}, I.~{Duran},
  R.~{Dressler}, N.~{Dzysiuk}, C.~{Eleftheriadis}, A.~{Ferrari}, K.~{Fraval},
  S.~{Ganesan}, A.~R. {Garc{\'\i}a}, G.~{Giubrone}, M.~B.
  {G{\'o}mez-Hornillos}, I.~F. {Gon{\c{c}}alves}, E.~{Gonz{\'a}lez-Romero},
  E.~{Griesmayer}, C.~{Guerrero}, F.~{Gunsing}, P.~{Gurusamy}, D.~G. {Jenkins},
  E.~{Jericha}, Y.~{Kadi}, F.~{K{\"a}ppeler}, D.~{Karadimos}, N.~{Kivel},
  P.~{Koehler}, M.~{Kokkoris}, G.~{Korschinek}, M.~{Krti{\v{c}}ka}, J.~{Kroll},
  C.~{Langer}, H.~{Leeb}, L.~S. {Leong}, R.~{Losito}, A.~{Manousos},
  J.~{Marganiec}, T.~{Mart{\'\i}nez}, P.~F. {Mastinu}, M.~{Mastromarco},
  M.~{Meaze}, E.~{Mendoza}, A.~{Mengoni}, P.~M. {Milazzo}, F.~{Mingrone},
  M.~{Mirea}, W.~{Mondelaers}, C.~{Paradela}, A.~{Pavlik}, J.~{Perkowski},
  M.~{Pignatari}, A.~{Plompen}, J.~{Praena}, J.~M. {Quesada}, T.~{Rauscher},
  R.~{Reifarth}, A.~{Riego}, F.~{Roman}, C.~{Rubbia}, R.~{Sarmento},
  P.~{Schillebeeckx}, S.~{Schmidt}, D.~{Schumann}, G.~{Tagliente}, J.~L.
  {Tain}, D.~{Tarr{\'\i}o}, L.~{Tassan-Got}, A.~{Tsinganis}, S.~{Valenta},
  G.~{Vannini}, V.~{Variale}, P.~{Vaz}, A.~{Ventura}, R.~{Versaci}, M.~J.
  {Vermeulen}, V.~{Vlachoudis}, R.~{Vlastou}, A.~{Wallner}, T.~{Ware},
  M.~{Weigand}, C.~{Wei{\ss}}, T.~J. {Wright}, and P.~{{\v{Z}}ugec}.
\newblock {Neutron Capture Cross Section of Unstable Ni63: Implications for
  Stellar Nucleosynthesis}.
\newblock {\em Phys. Rev. Lett.}, 110(2):022501, Jan 2013.

\bibitem{Moxon63}
M.~C. {Moxon} and E.~R. {Rae}.
\newblock {A gamma-ray detector for neutron capture cross-section
  measurements}.
\newblock {\em Nuclear Instruments and Methods}, 24:445--455, July 1963.

\bibitem{Macklin63a}
R.~L. {Macklin}, J.~H. {Gibbons}, and T.~{Inada}.
\newblock {Neutron capture cross sections near 30 keV using a Moxon-Rae
  detector}.
\newblock {\em Nuclear Physics}, 43:353--362, Jun 1963.

\bibitem{Macklin67}
R.~L. {Macklin} and J.~H. {Gibbons}.
\newblock {Capture-Cross-Section Studies for 30-220-keV Neutrons Using a New
  Technique}.
\newblock {\em Physical Review}, 159:1007--1012, July 1967.

\bibitem{Mathews86}
G.~J. {Mathews}, K.~{Takahashi}, R.~A. {Ward}, and W.~M. {Howard}.
\newblock {Stellar Technetium and Niobium Abundances as a Measure of the
  Lifetime of AGB Stars in the Third Dredge-up Phase}.
\newblock {\em The Astroph. Jour.}, 302:410, Mar 1986.

\bibitem{Corvi88}
F.~Corvi, A.~Prevignano, H.~Liskien, and P.B. Smith.
\newblock An experimental method for determining the total efficiency and the
  response function of a gamma-ray detector in the range 0.5–10 mev.
\newblock {\em Nuclear Instruments and Methods in Physics Research Section A:
  Accelerators, Spectrometers, Detectors and Associated Equipment}, 265(3):475
  -- 484, 1988.

\bibitem{Plag03}
R.~{Plag}, M.~{Heil}, F.~{K{\"a}ppeler}, P.~{Pavlopoulos}, R.~{Reifarth},
  K.~{Wisshak}, and {n TOF Collaboration}.
\newblock {An optimized C $_{6}$D $_{6}$ detector for studies of
  resonance-dominated (n,{$\gamma$}) cross-sections}.
\newblock {\em Nuclear Instruments and Methods in Physics Research A},
  496:425--436, January 2003.

\bibitem{Marrone04}
S.~{Marrone}, P.~F. {Mastinu}, U.~{Abbondanno}, R.~{Baccomi}, E.~Boscolo
  {Marchi}, N.~{Bustreo}, N.~{Colonna}, F.~{Gramegna}, M.~{Loriggiola},
  S.~{Marigo}, P.~M. {Milazzo}, C.~{Moreau}, M.~{Sacchetti}, G.~{Tagliente},
  R.~{Terlizzi}, G.~{Vannini}, G.~{Aerts}, E.~{Berthomieux}, D.~{Cano-Ott},
  P.~{Cennini}, C.~{Domingo-Pardo}, L.~{Ferrant}, E.~{Gonzalez-Romero},
  F.~{Gunsing}, M.~{Heil}, F.~{Kaeppeler}, T.~{Papaevangelou}, C.~{Paradela},
  P.~{Pavlopoulos}, L.~{Perrot}, R.~{Plag}, J.~L. {Tain}, H.~{Wendler}, and {n
  TOF Collaboration}.
\newblock {A low background neutron flux monitor for the n\_TOF facility at
  CERN}.
\newblock {\em Nuclear Instruments and Methods in Physics Research A},
  517(1-3):389--398, Jan 2004.

\bibitem{Tain04}
U.~{Abbondanno}, G.~{Aerts}, H.~{Alvarez}, S.~{Andriamonje}, A.~{Angelopoulos},
  P.~{Assimakopoulos}, C.~O. {Bacri}, G.~{Badurek}, P.~{Baumann}, F.~{Be{\v
  c}v{\'a}{\v r}}, H.~{Beer}, J.~{Benlliure}, B.~{Berthier}, E.~{Berthomieux},
  S.~{Boffi}, C.~{Borcea}, E.~{Boscolo-Marchi}, N.~{Bustreo}, P.~{Calvi{\~n}o},
  D.~{Cano-Ott}, R.~{Capote}, P.~{Carlson}, P.~{Cennini}, V.~{Chepel},
  E.~{Chiaveri}, C.~{Coceva}, N.~{Colonna}, G.~{Cortes}, D.~{Cortina},
  A.~{Couture}, J.~{Cox}, S.~{Dababneh}, M.~{Dahlfors}, S.~{David},
  R.~{Dolfini}, C.~{Domingo-Pardo}, I.~{Duran}, C.~{Eleftheriadis},
  M.~{Embid-Segura}, L.~{Ferrant}, A.~{Ferrari}, L.~{Ferreira-Lourenco},
  R.~{Ferreira-Marques}, H.~{Frais-Koelbl}, W.~I. {Furman}, Y.~{Giomataris},
  I.~F. {Goncalves}, E.~{Gonzalez-Romero}, A.~{Goverdovski}, F.~{Gramegna},
  E.~{Griesmayer}, F.~{Gunsing}, R.~{Haight}, M.~{Heil}, A.~{Herrera-Martinez},
  K.~G. {Ioannides}, N.~{Janeva}, E.~{Jericha}, F.~{K{\"a}ppeler}, Y.~{Kadi},
  D.~{Karamanis}, A.~{Kelic}, V.~{Ketlerov}, G.~{Kitis}, P.~E. {Koehler},
  V.~{Konovalov}, E.~{Kossionides}, V.~{Lacoste}, H.~{Leeb}, A.~{Lindote},
  M.~I. {Lopes}, M.~{Lozano}, S.~{Lukic}, S.~{Markov}, S.~{Marrone},
  J.~{Martinez-Val}, P.~{Mastinu}, A.~{Mengoni}, P.~M. {Milazzo}, E.~{Minguez},
  A.~{Molina-Coballes}, C.~{Moreau}, F.~{Neves}, H.~{Oberhummer}, S.~{O'Brien},
  J.~{Pancin}, T.~{Papaevangelou}, C.~{Paradela}, A.~{Pavlik},
  P.~{Pavlopoulos}, A.~{Perez-Parra}, J.~M. {Perlado}, L.~{Perrot},
  V.~{Peskov}, R.~{Plag}, A.~{Plompen}, A.~{Plukis}, A.~{Poch}, A.~{Policarpo},
  C.~{Pretel}, J.~M. {Quesada}, M.~{Radici}, S.~{Raman}, W.~{Rapp},
  T.~{Rauscher}, R.~{Reifarth}, F.~{Rejmund}, M.~{Rosetti}, C.~{Rubbia},
  G.~{Rudolf}, P.~{Rullhusen}, J.~{Salgado}, E.~{Savvidis}, J.~C. {Soares},
  C.~{Stephan}, G.~{Tagliente}, J.~L. {Tain}, C.~{Tapia}, L.~{Tassan-Got},
  L.~M.~N. {Tavora}, R.~{Terlizzi}, M.~{Terrani}, N.~{Tsangas}, G.~{Vannini},
  P.~{Vaz}, A.~{Ventura}, D.~{Villamarin-Fernandez}, M.~{Vincente-Vincente},
  V.~{Vlachoudis}, R.~{Vlastou}, F.~{Voss}, H.~{Wendler}, M.~{Wiescher},
  K.~{Wisshak}, L.~{Zanini}, and {n TOF Collaboration}.
\newblock {New experimental validation of the pulse height weighting technique
  for capture cross-section measurements}.
\newblock {\em Nuclear Instruments and Methods in Physics Research A},
  521:454--467, April 2004.

\bibitem{Pignatari10}
M.~{Pignatari}, R.~{Gallino}, M.~{Heil}, M.~{Wiescher}, F.~{K{\"a}ppeler},
  F.~{Herwig}, and S.~{Bisterzo}.
\newblock {The Weak s-Process in Massive Stars and its Dependence on the
  Neutron Capture Cross Sections}.
\newblock {\em The Astroph. Jour.}, 710(2):1557--1577, Feb 2010.

\bibitem{Zugec14}
P.~{{\v Z}ugec}, N.~{Colonna}, D.~{Bosnar}, S.~{Altstadt}, J.~{Andrzejewski},
  L.~{Audouin}, M.~{Barbagallo}, V.~{B{\'e}cares}, F.~{Be{\v c}v{\'a}{\v r}},
  F.~{Belloni}, E.~{Berthoumieux}, J.~{Billowes}, V.~{Boccone}, M.~{Brugger},
  M.~{Calviani}, F.~{Calvi{\~n}o}, D.~{Cano-Ott}, C.~{Carrapi{\c c}o},
  F.~{Cerutti}, E.~{Chiaveri}, M.~{Chin}, G.~{Cort{\'e}s}, M.~A.
  {Cort{\'e}s-Giraldo}, M.~{Diakaki}, C.~{Domingo-Pardo}, R.~{Dressler},
  I.~{Duran}, N.~{Dzysiuk}, C.~{Eleftheriadis}, A.~{Ferrari}, K.~{Fraval},
  S.~{Ganesan}, A.~R. {Garc{\'{\i}}a}, G.~{Giubrone}, M.~B.
  {G{\'o}mez-Hornillos}, I.~F. {Gon{\c c}alves}, E.~{Gonz{\'a}lez-Romero},
  E.~{Griesmayer}, C.~{Guerrero}, F.~{Gunsing}, P.~{Gurusamy}, S.~{Heinitz},
  D.~G. {Jenkins}, E.~{Jericha}, Y.~{Kadi}, F.~{K{\"a}ppeler}, D.~{Karadimos},
  N.~{Kivel}, P.~{Koehler}, M.~{Kokkoris}, M.~{Krti{\v c}ka}, J.~{Kroll},
  C.~{Langer}, C.~{Lederer}, H.~{Leeb}, L.~S. {Leong}, S.~{Lo Meo},
  R.~{Losito}, A.~{Manousos}, J.~{Marganiec}, T.~{Mart{\'{\i}}nez},
  C.~{Massimi}, P.~F. {Mastinu}, M.~{Mastromarco}, M.~{Meaze}, E.~{Mendoza},
  A.~{Mengoni}, P.~M. {Milazzo}, F.~{Mingrone}, M.~{Mirea}, W.~{Mondalaers},
  C.~{Paradela}, A.~{Pavlik}, J.~{Perkowski}, A.~{Plompen}, J.~{Praena}, J.~M.
  {Quesada}, T.~{Rauscher}, R.~{Reifarth}, A.~{Riego}, F.~{Roman}, C.~{Rubbia},
  R.~{Sarmento}, A.~{Saxena}, P.~{Schillebeeckx}, S.~{Schmidt}, D.~{Schumann},
  G.~{Tagliente}, J.~L. {Tain}, D.~{Tarr{\'{\i}}o}, L.~{Tassan-Got},
  A.~{Tsinganis}, S.~{Valenta}, G.~{Vannini}, V.~{Variale}, P.~{Vaz},
  A.~{Ventura}, R.~{Versaci}, M.~J. {Vermeulen}, V.~{Vlachoudis}, R.~{Vlastou},
  A.~{Wallner}, T.~{Ware}, M.~{Weigand}, C.~{Wei{\ss}}, and T.~{Wright}.
\newblock {GEANT4 simulation of the neutron background of the C$_{6}$D$_{6}$
  set-up for capture studies at n\_TOF}.
\newblock {\em Nuclear Instruments and Methods in Physics Research A},
  760:57--67, October 2014.

\bibitem{Tagliente13}
G.~{Tagliente}, P.~M. {Milazzo}, K.~{Fujii}, U.~{Abbondanno}, G.~{Aerts},
  H.~{{\'A}lvarez}, F.~{Alvarez-Velarde}, S.~{Andriamonje}, J.~{Andrzejewski},
  L.~{Audouin}, G.~{Badurek}, P.~{Baumann}, F.~{Be{\v{c}}v{\'a}{\v{r}}},
  F.~{Belloni}, E.~{Berthoumieux}, F.~{Calvi{\~n}o}, M.~{Calviani},
  D.~{Cano-Ott}, R.~{Capote}, C.~{Carrapi{\c{c}}o}, P.~{Cennini}, V.~{Chepel},
  E.~{Chiaveri}, N.~{Colonna}, G.~{Cortes}, A.~{Couture}, M.~{Dahlfors},
  S.~{David}, I.~{Dillmann}, C.~{Domingo-Pardo}, W.~{Dridi}, I.~{Duran},
  C.~{Eleftheriadis}, M.~{Embid-Segura}, A.~{Ferrari}, R.~{Ferreira-Marques},
  W.~{Furman}, I.~{Goncalves}, E.~{Gonzalez-Romero}, F.~{Gramegna},
  C.~{Guerrero}, F.~{Gunsing}, B.~{Haas}, R.~{Haight}, M.~{Heil},
  A.~{Herrera-Martinez}, E.~{Jericha}, F.~{K{\"a}ppeler}, Y.~{Kadi},
  D.~{Karadimos}, D.~{Karamanis}, M.~{Kerveno}, E.~{Kossionides},
  M.~{Krti{\v{c}}ka}, C.~{Lamboudis}, H.~{Leeb}, A.~{Lindote}, I.~{Lopes},
  S.~{Lukic}, J.~{Marganiec}, S.~{Marrone}, T.~{Mart{\'\i}nez}, C.~{Massimi},
  P.~{Mastinu}, A.~{Mengoni}, C.~{Moreau}, M.~{Mosconi}, F.~{Neves},
  H.~{Oberhummer}, S.~{O'Brien}, C.~{Papachristodoulou}, C.~{Papadopoulos},
  C.~{Paradela}, N.~{Patronis}, A.~{Pavlik}, P.~{Pavlopoulos}, L.~{Perrot},
  M.~T. {Pigni}, R.~{Plag}, A.~{Plompen}, A.~{Plukis}, A.~{Poch}, J.~{Praena},
  C.~{Pretel}, J.~{Quesada}, R.~{Reifarth}, M.~{Rosetti}, C.~{Rubbia},
  G.~{Rudolf}, P.~{Rullhusen}, J.~{Salgado}, C.~{Santos}, L.~{Sarchiapone},
  I.~{Savvidis}, C.~{Stephan}, J.~L. {Tain}, L.~{Tassan-Got}, L.~{Tavora},
  R.~{Terlizzi}, G.~{Vannini}, P.~{Vaz}, A.~{Ventura}, D.~{Villamarin}, M.~C.
  {Vincente}, V.~{Vlachoudis}, R.~{Vlastou}, F.~{Voss}, S.~{Walter},
  M.~{Wiescher}, and K.~{Wisshak}.
\newblock {The $^{93}$Zr(n,{\ensuremath{\gamma}}) reaction up to 8 keV neutron
  energy}.
\newblock {\em Phys. Rev. C}, 87(1):014622, Jan 2013.

\bibitem{Neyskens15}
P.~{Neyskens}, S.~{van Eck}, A.~{Jorissen}, S.~{Goriely}, L.~{Siess}, and
  B.~{Plez}.
\newblock {The temperature and chronology of heavy-element synthesis in
  low-mass stars}.
\newblock {\em Nature}, 517(7533):174--176, Jan 2015.

\bibitem{Perez16}
D.~L. {P{\'e}rez Mag{\'a}n}, L.~{Caballero}, C.~{Domingo-Pardo},
  J.~{Agramunt-Ros}, F.~{Albiol}, A.~{Casanovas}, A.~{Gonz{\'a}lez},
  C.~{Guerrero}, J.~{Lerendegui-Marco}, and A.~{Tarife{\~n}o-Saldivia}.
\newblock {First tests of the applicability of {$\gamma$}-ray imaging for
  background discrimination in time-of-flight neutron capture measurements}.
\newblock {\em Nuclear Instruments and Methods in Physics Research A},
  823:107--119, July 2016.

\bibitem{Schoenfelder73}
V.~{Sch\"onfelder}, A.~{Hirner}, and K.~{Schneider}.
\newblock {A telescope for soft gamma ray astronomy}.
\newblock {\em Nuclear Instruments and Methods}, 107:385--394, 1973.

\bibitem{Domingo16}
C.~{Domingo-Pardo}.
\newblock {i-TED: A novel concept for high-sensitivity (n,{$\gamma$})
  cross-section measurements}.
\newblock {\em Nuclear Instruments and Methods in Physics Research A},
  825:78--86, July 2016.

\bibitem{hymns}
{High-sensitivitY Measurements of key stellar Nucleo-Synthesis reactions
  (HYMNS), ERC-Consolidator Grant Agreement No. 681740, PI: C. Domingo Pardo}.

\bibitem{Babiano19b}
V.~{Babiano}, L.~{Caballero}, D.~{Calvo}, I.~{Ladarescu}, S.~Mira {Prats}, and
  C.~{Domingo-Pardo}.
\newblock {First i-TED demonstrator: a Compton imager with Dynamic Electronic
  Collimation}.
\newblock {\em arXiv e-prints}, page arXiv:1908.08533, Aug 2019.

\bibitem{Mihailescu07}
L.~Mihailescu, K.M. Vetter, M.T. Burks, E.L. Hull, and W.W. Craig.
\newblock Speir: A ge compton camera.
\newblock {\em Nuclear Instruments and Methods in Physics Research Section A:
  Accelerators, Spectrometers, Detectors and Associated Equipment}, 570(1):89
  -- 100, 2007.

\bibitem{Kishimoto17}
A.~{Kishimoto}, J.~{Kataoka}, A.~{Koide}, K.~{Sueoka}, Y.~{Iwamoto}, T.~{Taya},
  and S.~{Ohsuka}.
\newblock {Development of a compact scintillator-based high-resolution Compton
  camera for molecular imaging}.
\newblock {\em Nuclear Instruments and Methods in Physics Research A},
  845:656--659, Feb 2017.

\bibitem{Nagao18}
Yuto {Nagao}, Mitsutaka {Yamaguchi}, Naoki {Kawachi}, and Hiroshi {Watabe}.
\newblock {Development of a cost-effective Compton camera using a positron
  emission tomography data acquisition system}.
\newblock {\em Nuclear Instruments and Methods in Physics Research A},
  912:20--23, Dec 2018.

\bibitem{Olleros18}
P.~{Olleros}, L.~{Caballero}, C.~{Domingo-Pardo}, V.~{Babiano}, I.~{Ladarescu},
  D.~{Calvo}, P.~{Gramage}, E.~{Nacher}, J.~L. {Tain}, and A.~{Tolosa}.
\newblock {On the performance of large monolithic LaCl$_{3}$(Ce) crystals
  coupled to pixelated silicon photosensors}.
\newblock {\em Journal of Instrumentation}, 13:P03014, March 2018.

\bibitem{Li10}
Z.~{Li}, M.~{Wedrowski}, P.~{Bruyndonckx}, and G.~{Vandersteen}.
\newblock {Nonlinear least-squares modeling of 3D interaction position in a
  monolithic scintillator block}.
\newblock {\em Physics in Medicine and Biology}, 55:6515--6532, November 2010.

\bibitem{Babiano19}
V.~{Babiano}, L.~{Caballero}, D.~{Calvo}, I.~{Ladarescu}, P.~{Olleros}, and
  C.~{Domingo-Pardo}.
\newblock {{\ensuremath{\gamma}}-Ray position reconstruction in large
  monolithic LaCl$_{3}$(Ce) crystals with SiPM readout}.
\newblock {\em Nuclear Instruments and Methods in Physics Research A},
  931:1--22, Jul 2019.

\bibitem{petsys}
A.~{Di Francesco}, R.~{Bugalho}, L.~{Oliveira}, A.~{Rivetti}, M.~{Rolo}, J.~C.
  {Silva}, and J.~{Varela}.
\newblock {TOFPET 2: A high-performance circuit for PET time-of-flight}.
\newblock {\em Nuclear Instruments and Methods in Physics Research A},
  824:194--195, July 2016.

\end{thebibliography}


\end{document}